\documentclass[manuscript,screen,acmlarge]{acmart}
\usepackage{libertine}
\usepackage{xspace}

\usepackage{balance}       
\usepackage{graphics}      
\usepackage[T1]{fontenc}   
\usepackage{color}
\usepackage{booktabs}
\usepackage{textcomp}
\usepackage{multirow}

\usepackage{microtype}        
\usepackage{ccicons}          

\newcommand{\tool}{{\texttt{Voicify}}\xspace}

\def\plainkeywords{Authors' choice; of terms; separated; by
  semicolons; include commas, within terms only; this section is required.}

\makeatletter

 \makeatletter

\def\pprw{8.5in}
\def\pprh{11in}

\setcopyright{acmlicensed}
\acmJournal{IMWUT}
\acmYear{2023} \acmVolume{7} \acmNumber{1} \acmArticle{44} \acmMonth{3} \acmPrice{15.00}\acmDOI{10.1145/3581998}

\begin{document}

\title{Voicify Your UI: Towards Android App Control with Voice Commands}
\author{Minh Duc Vu}
\email{dustin.vu@monash.edu}
\orcid{https://orcid.org/0000-0002-4798-8701}
\author{Han Wang}
\email{han.wang@monash.edu}
\orcid{https://orcid.org/0000-0001-7862-6677}
\author{Zhuang Li}
\email{zhuang.li@monash.edu}
\orcid{https://orcid.org/0000-0002-9808-9992}
\authornote{Minh Duc Vu, Han Wang, and Zhuang Li contributed equally.}
\affiliation{%
  \institution{Monash University}
  \city{Melbourne}
  \country{Australia}}

\author{Gholamreza Haffari}
\affiliation{%
  \institution{Monash University}
  \city{Melbourne}
  \country{Australia}}
\email{ gholamreza.haffari@monash.edu}
\orcid{https://orcid.org/0000-0001-7326-8380}

\author{Zhenchang Xing}
  \affiliation{%
  \institution{CSIRO's Data61 \& Australian National University}
  \city{Canberra}
  \country{Australia}}
\email{zhenchang.xing@data61.csiro.au}
\orcid{https://orcid.org/0000-0001-7663-1421}

\author{Chunyang Chen}
\affiliation{%
  \institution{Monash University}
  \city{Melbourne}
  \country{Australia}}
\email{chunyang.chen@monash.edu}
\orcid{https://orcid.org/0000-0003-2011-9618}
\authornote{Chunyang Chen is the corresponding author.}
\renewcommand{\shortauthors}{Vu, Wang, and Li, et al.}

\begin{abstract}
Nowadays, voice assistants help users complete tasks on the smartphone with voice commands, replacing traditional touchscreen interactions when such interactions are inhibited. However, the usability of those tools remains moderate due to the problems in understanding rich language variations in human commands, along with efficiency and comprehensibility issues. Therefore, we introduce \tool, an Android virtual assistant that allows users to interact with on-screen elements in mobile apps through voice commands. Using a novel deep learning command parser, \tool interprets human verbal input and performs matching with UI elements. In addition, the tool can directly open a specific feature from installed applications by fetching application code information to explore the set of in-app components. Our command parser achieved 90\% accuracy on the human command dataset. Furthermore, the direct feature invocation module achieves better feature coverage in comparison to Google Assistant. The user study demonstrates the usefulness of \tool in real-world scenarios.
\end{abstract}



\maketitle


\section{Introduction}
\label{sec:introduction}

\begin{figure}
\centering
\includegraphics[width=\textwidth]{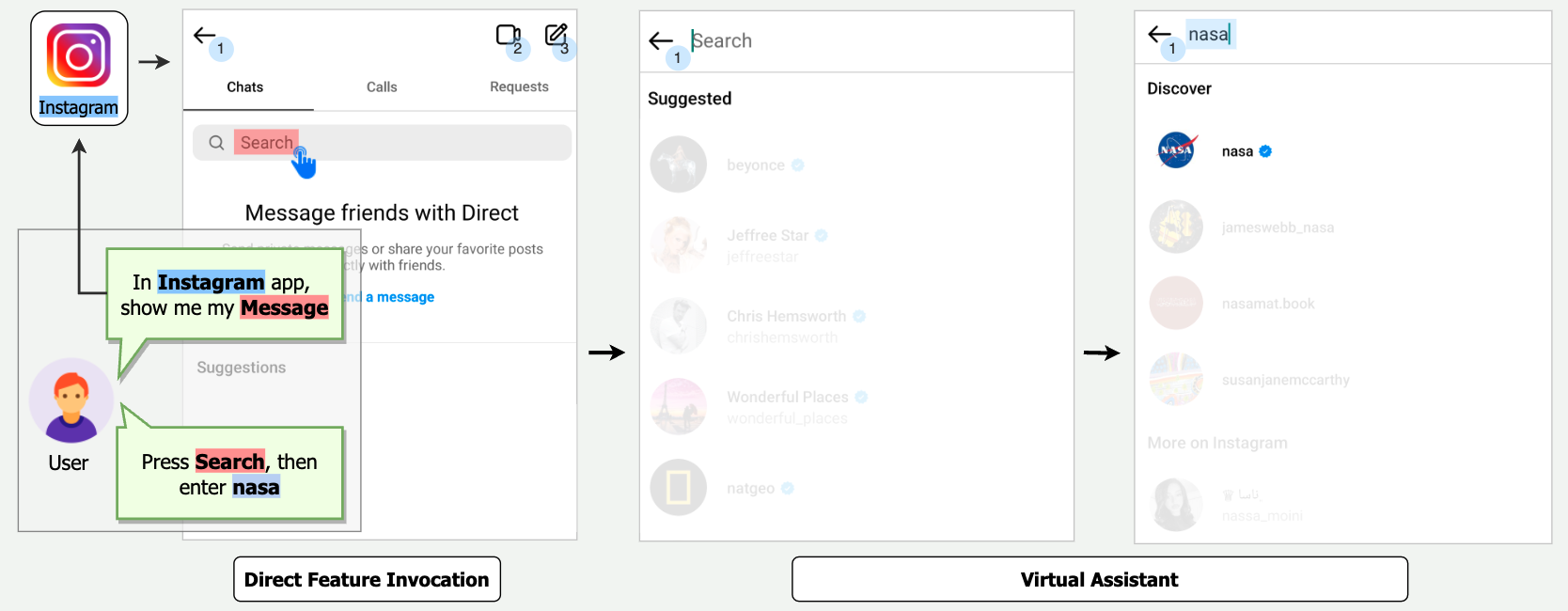}
\caption{Voicify provides an intuitive way to interact with Android devices, as users can directly open a feature from other applications and interact with UI elements using voice commands. We show real-world usage of searching for nasa in Instagram's messaging feature.}
\label{fig:thumbnail_image}
\end{figure}

Smart mobile devices have revolutionised user experience, in which touchscreens are becoming more and more popular. In normal usage scenarios, interacting with a touchscreen is handy and intuitive, however, it is a tedious process for users whose hands are occupied (e.g., cooking, holding a baby or typing) and those with temporary hand injuries or permanent motor impairment ~\cite{zhong_raman_burkhardt_biadsy_bigham_2014}. Under those circumstances, attempting to perform physical interactions on the phone such as tapping or scrolling is extremely inconvenient and imprecise. On the other hand, researchers have pointed out other problems associated with physical mobile interaction. Despite the increasing screen size~\cite{kim2011effects}, interactions become hindered because of the reduction in thumb movement coverage~\cite{xiong2016effects}, requiring users' efforts to reach far-part UI elements. On smaller screen sizes, misclicking regularly happens due to clustered elements and complex usage scenarios~\cite{hu2018user}. Therefore, users envisage an alternative way to interact with smart devices that help them to overcome these aforementioned issues.

Advanced technologies have fueled the development of voice assistants, which provide an innovative solution to the described problems of physical touches, working as a hands-free input method under many circumstances. This technology changes how individuals interact with mobile devices to achieve daily tasks. The most well-known voice assistants include Alexa, Siri and Cortana, which are integrated into various smart devices across multiple platforms~\cite{hoy2018alexa}. In Android, Google Assistant~\cite{google_assistant} and Google Voice Access~\cite{hume_2020} are the most frequently used voice assistants. The distinguishable functionality between these two assistants is that while Voice Access offers full control for on-screen interactions but cannot reach an in-app feature directly, Google Assistant supports users to directly open a specific feature from other applications. Apart from these aforementioned Google systems that utilize voice assistant technology, researchers have proposed other systems to leverage the usage of voice interactions with Android devices~\cite{zhong_raman_burkhardt_biadsy_bigham_2014,bhalerao2017smart,liu2015voice}.

However, there is room for further research as some key functionalities are still missing or can be extended. Many research only targeted a fixed set of tasks including messaging, calling or changing the device's settings using voice commands~\cite{liu2015voice,kulhalli2017personal,khan2018blindsense}, they do not offer full control over one’s device, hence cannot replace touch gestures such as tapping or swiping. Voice Access addressed this issue by providing a system that allows users to fully control their Android devices using voice commands. However, the system has a limited number of features and users have reported reliability issues within the system~\cite{voice_access_issue}. Our experiment also shows that Voice Access requires a steep learning curve due to the complexity of command syntax. On the other hand, the ability to directly open a given feature from other applications has been overlooked or pretermitted by smartphone voice assistants. Many useful features from an application are located under the deep hierarchical navigation ~\cite{foster2011barrel}, requiring several steps to reach from the home screen. Therefore, allowing users to reach the requested application screen is highly usable in a real-world context. Google Assistant has a fixed list of application features that can be invoked directly (namely App Action~\cite{google_assistant}), which were registered by app developers. Because the registration process is tedious and requires extra effort, many apps omit the required declarations, hence their functionalities cannot be invoked by Google Assistant~\cite{arsan2021app}. Another issue with the aforementioned systems is that they are not open-sourced, which prevents other developers from further researching and integrating with the systems.

This paper introduces \tool (see Fig.~\ref{fig:thumbnail_image}), an application with novel approaches to enhance the usability of voice control in Android devices. The system works as a background service that interprets user's voice commands and executes them. We proposed a semantic parser that can bridge the gap between natural language from human commands and interactive tasks on mobile devices. \tool allows users to perform daily physical interactions with mobile devices such as scrolling, tapping or inputting text completely using voice commands. We achieve the mentioned functionalities by mapping the structured actions from the semantic parser to the collected on-screen UI elements. In addition, \tool can invoke a specific feature from an installed application directly without any extra integration work for application developers. We achieved this by fetching available components from applications that are installed locally on the user's device. Furthermore, users can give several commands within one utterance, the application will execute each command sequentially.

We validated the technical contributions of \tool by evaluating the command parser and the direct invocation module. The command parser achieved over 90\% accuracy on the testing data extracted from the AndroidHowTo dataset~\cite{seq2act}, which consisted of short-form human commands to perform specific tasks on Android devices. We then analysed the ability to directly open a feature from other applications in comparison to Google Assistant as a baseline. The result showed that \tool has 76.9\% feature coverage on the dataset of 119 in-app features extracted from 30 well-known applications, which was greater than 47\% of Google Assistant. Lastly, we also conducted a user experiment to evaluate the system's usability in real-world use cases. Participants were asked to complete given tasks using voice commands and provide feedback for \tool compared to Voice Access as the baseline. The experimental result showed that \tool is well-integrated and easy to use, which helped users achieve better performance on given tasks. 

To summarize, the contributions of this paper include:
\begin{itemize}
    \item A highly usable voice control and navigation system, \tool, that can recognise user vocal commands, associate them with app features (for direct feature invocations) or on-screen elements (for basic interactions) based on the pre-extracted and run-time application data, and execute them accordingly on Android devices. 
      \item A deep learning-based human command parser that can correctly map a user's command into executable actions in the context of mobile apps. 
    \item Our experiments and user studies show that \tool is easy to learn and use, and requires a minimal cognitive load to achieve better performance than the existing product when using real-world apps.
    \item \tool\footnote{\url{https://github.com/vuminhduc796/Voicify}} is open-sourced so that anyone can use and continue to improve the system. 
\end{itemize}

\section{Background \& Related works}
\label{sec:relatedWorks}


\subsection{Natural Language Understanding in Voice Assistants}
 Natural Language Understanding (NLU) systems aim to interpret and process the user’s speech input. In recent years, research has been conducted in the space of natural language interpretation and mapping human instructions to user interface actions. Seq2Act model has been introduced to extract actions (such as open, click and navigate) and object targets from user instructions and associate them with mobile UI elements to support executable action sequences~\cite{seq2act}. The technology is not only limited to usages within the smartphone but also extended to other smart systems~\cite{krishna2012zigbee,bai2022research} and home appliances~\cite{park2018low}.
 
 In Android, JustSpeak was launched with advanced command-chaining recognition and extended Google Automatic Speech Recognition (ASR) by its utterance parsing technique~\cite{zhong_raman_burkhardt_biadsy_bigham_2014}. The predefined set of commands from JustSpeak is later extended by Smart Voice Assistant to handle calls and SMS using voice commands~\cite{bhalerao2017smart}. In recent years, many technologies have been proposed to make NLP components more approachable, hence improving the applicability in different Android systems. Arsan et al. used Dialogflow as a conversational agent, providing a platform that can extract the intent from user utterances~\cite{arsan2021app}. In addition, by utilizing the pre-built Almond language model, DoThisHere provides a voice controlling system to get and set UI contents in Android \cite{yang2020dothishere}. However, the inflexible nature of pre-built language models from Almond and Dialogflow limits the capability to be extended and fully support our use cases. Therefore, \tool introduced an extensible parser using deep learning approach to cater for the flexibility in human commands. We propose the model as a reusable solution to improve communication between humans and technological devices.

\subsection{Android UI Semantics \& On-screen Interaction}

A core challenge of developing a system-level assistive tool is processing and interacting with on-screen UI semantics in run-time. Android has introduced Accessibility Service API in Android 1.6, allowing the developer to create accessibility tools that assist different types of impairments~\cite{accessibility_service}. This API allows Android developers to access additional accessibility metadata and the displaying content of the UI window and supports on-screen interactions with the device. Accessibility Service API has brought forth opportunities that propel the development of assistive and automation tools for Android~\cite{bhalerao2017smart, xie_ni_liu_cao_chen_2021,salehnamadi2021latte}. Those tools perform data collection from user's devices for analysis or perform sequential actions by matching the input with extracted on-screen textual data.

One of the earliest applications to leverage this technology was JustSpeak~\cite{zhong_raman_burkhardt_biadsy_bigham_2014}. The solution supports on-screen interactions via voice commands such as tapping on-screen elements, based on the collected accessibility metadata. Similar work was conducted in Weber et al.’s ``VoiceNavigator” application in 2016, which was the centre of their study on improving the visibility and learnability of mobile voice user interface applications~\cite{corbett_weber_2016}. Since then, Google has released Voice Access~\cite{hume_2020}, which aims to assist people with temporary injury or motor impairment with basic navigation and controlling the current screen. While Voice Access is the most popular voice control in Android, its low usability negatively affects the user experience, according to the user reviews on Play Store listing~\cite{voice_access_review}. Using a wide range of advanced features from Accessibility API, we propose \tool to improve upon existing solutions and enhance the intuitiveness of Android assistive tool.

\subsection{Android Components \& Direct Feature Invocation}

 In Android, the transition between different activities (representing UI screens) happens when the user navigates to another screen. An activity can invoke another activity by sending an Intent messaging object to the Android OS, which is then used to retrieve the corresponding intent filter and eventually navigate to the destination activity. Although sending intent objects is powerful as it can request features from any installed applications~\cite{alhanahnah2020dina}, intent-based invocation has not been commonly applied for direct component communication. One of the recent works in this area applied intent-based communication for the component invocation to create application shortcuts~\cite{arsan2021app}. While their approach requires static analysis of a given dataset to generate a pre-defined database of shortcuts, \tool retrieves the data from on-device applications, which avoids the limitation of dataset coverage. On the other hand, Google has provided Google Assistant as a part of Android OS, which can directly open a feature from applications via intent invocation. However, the platform has limited coverage of features since it does not automatically compatible with Android applications. External developers are required to manually declare the mapping between their existing features as intents and corresponding voice command to trigger them and must frequently update the declarations. \tool reads such data directly from the application files, which does not need extra effort from developers to comply with the platform. 

Recently, Deep link has been introduced by Android, which allows the developer to assign a unique resource identifier (URI) to specific a feature within the app~\cite{deep_links}. Several researchers have been improving the usability and performance of the Android native deep link system to enhance the ability to directly open a feature within an app. A record and replay implementation was proposed in the uLink framework to improve the default deep linking system, which provides a lightweight, universal solution with reduced developer efforts~\cite{azim_riva_nath_2016}. Similarly, DroidLink proposed a solution that uses a model to analyze the transition of activities within the provided app to build shortcuts to different UI elements~\cite{ma2016droidlink}. While the mentioned approaches attempted to replicate the behaviour of Android deep links, the authors have mentioned the reliability issues and limitations in performing some actions. Therefore, \tool utilizes deep links that have been created by developers for their applications to directly open its features.

\section{Motivation}
\label{sec:study}

\subsection{Effects of Learnability and Ease of Operation on User Experience for Voice Controlling System}
In real-world usage, voice control systems are mostly used in situations where users are busy with other activities~\cite{zhong_raman_burkhardt_biadsy_bigham_2014,interruptibility}. It is deduced that when users are focusing on their primary activity, voice control is the common choice to interact with mobile devices to perform a secondary activity. Therefore, users would not be able to fully concentrate on their voice-controlling tasks. In addition, researchers have pointed out that learnability and discoverability can strongly affect user experience with voice control system~\cite{learnability_voice}. Therefore, it is integral to aim for ease of operation and minimal mental demand to improve the usability of \tool.

\subsection{Usefulness of Providing In-App Feature Shortcuts}
Due to the small screen size, mobile apps often contain multiple screens. Therefore, it is time-consuming and complicated to reach the final destination screen for users to perform the task, given that the entire process is done via voice commands. Therefore, providing shortcuts to specific screens inside applications is useful. In Android, the idea has been implemented by Google Assistant to directly open a feature from a given application. However, Google Assistant has low feature coverage due to its implementation~\cite{arsan2021app}, as well as lacking a method that supports users to fully complete user's intended task via voice command to cater for hands-free usages. Hence, we proposed \tool to open specific features inside the applications while avoiding Google Assistant's shortcomings.

\subsection{Robustness to the Rich Linguistic Variations in the Human Command Utterances}
One major issue that caused the inaccuracy of the voice system was the rich linguistic variations in human utterances. Users might describe one action or target using multiple word choices. For example, different synonyms of ``tap'' such as ``click'' or ``press'' are spoken by the users to perform a tap on the screen. In addition, the application name such as ``Gmail'', might be mentioned as ``Gmail'' or ``Google mail''. The complexity of multiple variations in word choice within a single command caused NLP systems to misunderstand the user's intention in their command utterances~\cite{wijeratne2019natural}, therefore, requiring a better approach.

\subsection{Design Goals}
Based on the motivation in the previous section, we propose the design goals to improve the usability of existing tools: 
\begin{itemize}

  \item \textbf{Comprehensibility:} Reduce the learning curve of the system and require minimal cognitive load from users to operate the tool.
  \item \textbf{Efficiency:} Provide direct access to in-app features without navigating through multiple screens using voice commands to reduce the time to complete a certain activity
  \item \textbf{Robustness:} Understand the user's intention in human speech utterances with rich linguistic variations to avoid misinterpretation of user commands.

\end{itemize}

\section{The Voicify System}

\begin{figure*}[!h]
\centering
\includegraphics[width=\textwidth]{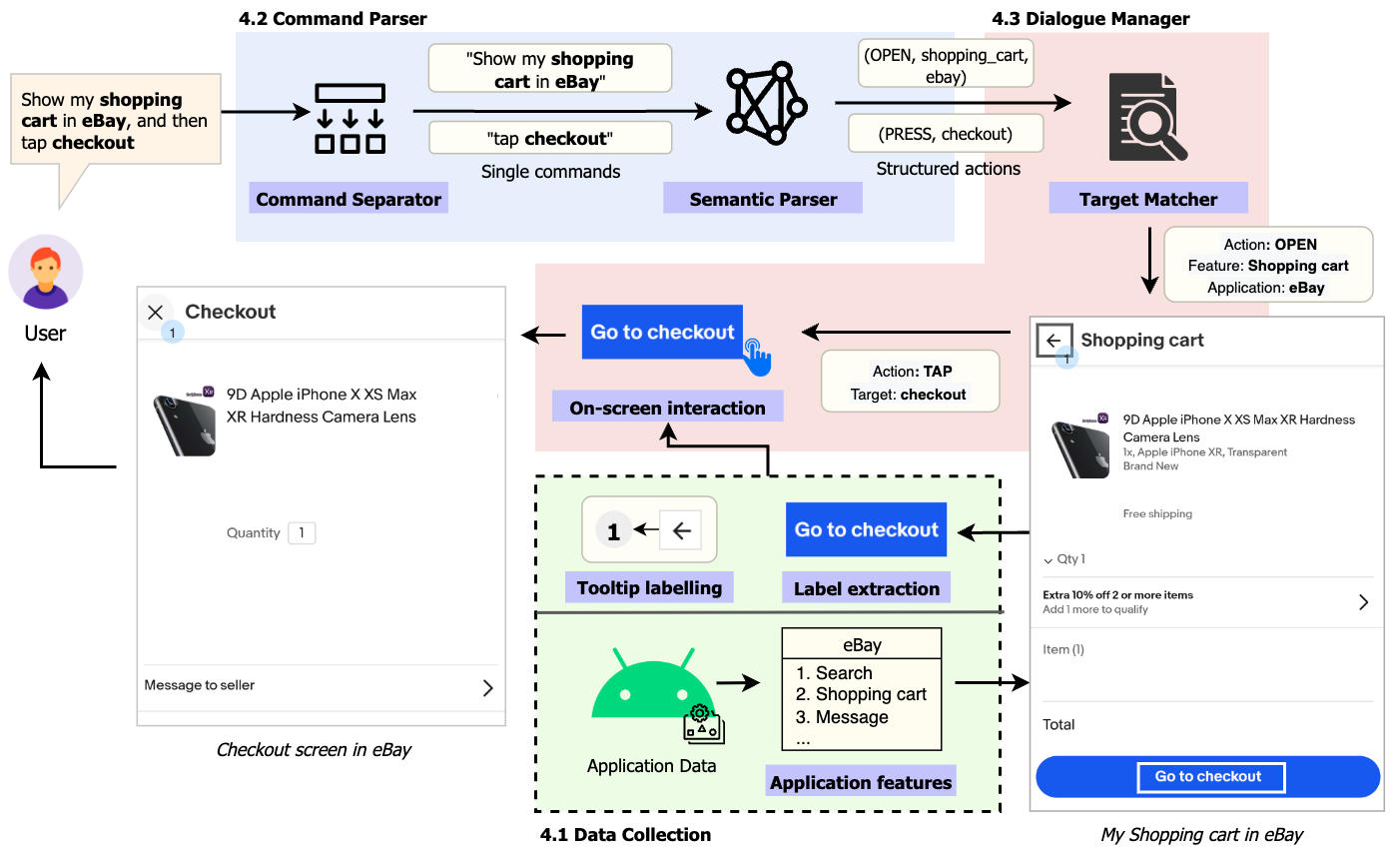}
\caption{An example use case in eBay application to describe the overall system design of \tool. Upon receiving user commands, the 4.2) Command Parser separates them into single commands and parses them into structured actions,
which are used by the 4.3) Dialogue Manager to perform target matching and sequentially execute each action. The 4.1) Data Collection module works in the background to provide static application components and interpret dynamic on-screen UI semantics. }
\label{fig:overview}
\end{figure*}

\label{sec:voicify}
     Based on the design implications, we propose \tool to leverage the voice interaction between users and Android devices. Our system provides a background service, which continuously listens to users’ commands and executes the action. We solved the problem of misinterpreting user commands by proposing a deep learning command parser that caters for the rich linguistic variation in human speech. \tool allows users to directly interact with UI elements using on-screen labels and numbering tooltips, hence improving the comprehensibility of the system. In addition, by processing the on-device application files, the system improves voice control efficiency by directly opening a feature that matches user requests. Our approach overcomes the limitation in the feature coverage of Google Assistant, minimizing developer efforts to comply with the Google platform. The overall system design is demonstrated in Fig.~\ref{fig:overview}, which consists of three modules: Data Collection (Section~\ref{sec:dataExtraction}), Command Parser (Section~\ref{sec:commandParser}) and Dialogue Manager (Section~\ref{sec:actionExecution}).

\subsection{Data Collection}
\label{sec:dataExtraction}
     \tool efficiently explores and processes on-device data in the Android environment. To provide contextual data for executing actions, we collect both (i) static data about application components for direct invocation and (ii) dynamic data from the device's screen to identify on-screen UI elements and perform interactions.
    
    \begin{figure*}[!h]
    \centering
    \includegraphics[width=0.88\textwidth]{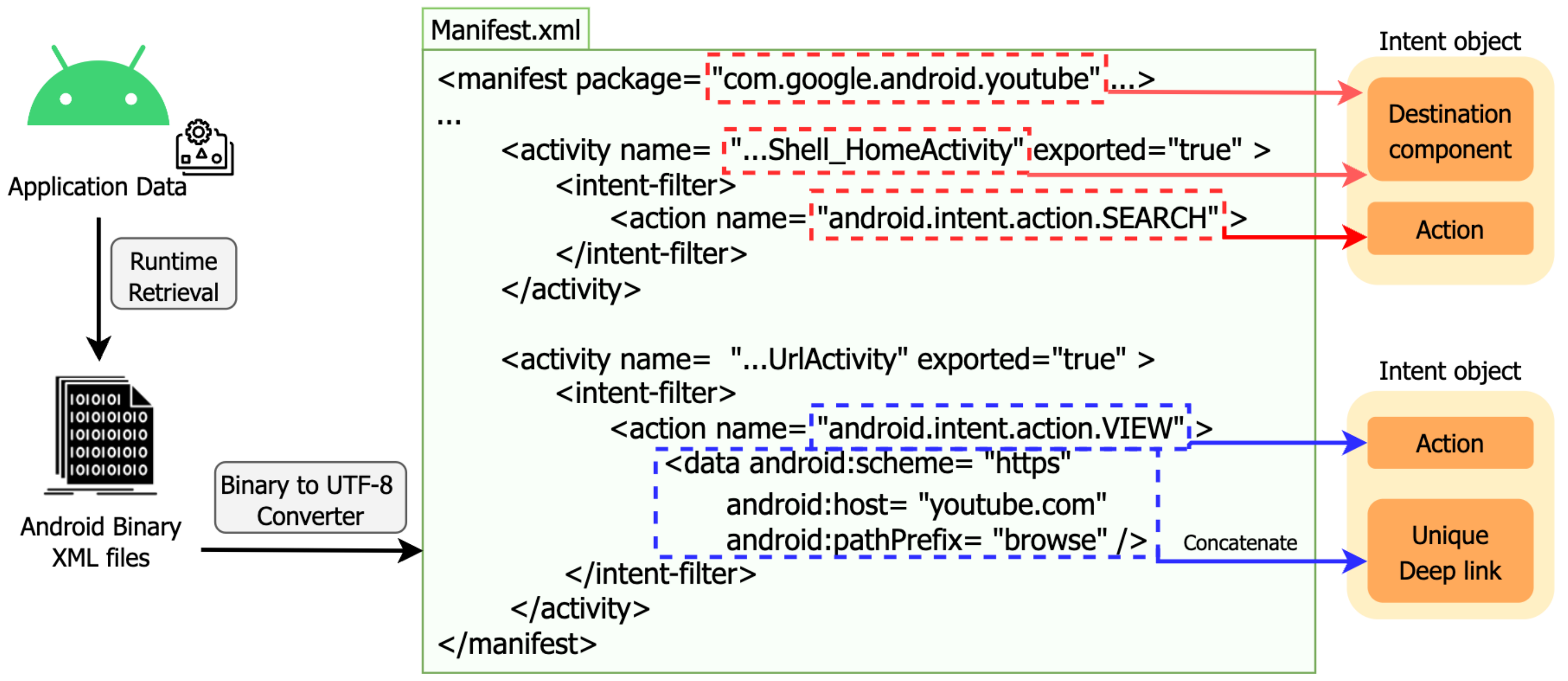}
    \caption{The pipeline to produce the intent object for each activity using the raw Application data. We used the snippet of the Manifest.xml file from YouTube application to highlight deep links and app components.}
    \label{fig:extraction}
    \end{figure*}
    
    \subsubsection{Application Components Exploration} 
    To provide direct invocation for different app components, \tool fetches the system files from locally installed applications that specifies the list of components provided by the app. Thus, all installed applications will be compatible with \tool without any additional integration from the other app developers, resulting in higher application feature coverage. In addition, local retrieval will prevent application version mismatch when the locally installed version is not up-to-date with the released version from the developers, which was a reported issue that caused Google Assistant to malfunction~\cite{xaif_bam_ram_ajinkya_patel_kerwin_2022}.

    Our system uses explicit intents for the direct invocation process, requiring an activity name or a deep link to determine the target component. These data can be found inside the \emph{Manifest.xml} file within the application source code. Therefore, \tool locates and retrieves the \emph{Manifest.xml} file from the on-device application's directories, as shown in Fig.~\ref{fig:extraction}. The data inside application directories are encoded in Android-specific binary XML format, allocated into chunks with little-endian byte order by default~\cite{binary_xml}. Therefore, we introduced a binary decoder to convert the Manifest file for each installed application into UTF-8-character representation.
    
     \tool retrieves the intent filter for each exported component, which acts as the access point to open that component. Using the attributes from retrieved intent filters, we construct corresponding intents that can directly invoke that intent filter. In Fig.~\ref{fig:extraction}, we illustrate the representation of deep links (blue boxes) and exported components (red boxes) in the manifest file for the YouTube application. Since a deep link is a unique resource identifier, an intent that attaches a deep link can directly invoke an app component. The system retrieves it by concatenating the host, scheme and path prefix in the \emph{<data>} tag. To open application components that are not integrated with deep links, we use the package name and the activity name to identify the destination component. \tool utilizes the declared action name such as SEARCH and VIEW to improve the granularity of feature exploration. All extracted intents will be used for direct invocation process in Section~\ref{sec:actionExecution}.
    
    \subsubsection{UI Semantic Retrieval}

    In \tool, UI semantic retrieval is an essential module that allows the system to identify the UI elements for interactions. Similar to Kalysch et al.~\cite{kalysch2018android}, we utilized Android Accessibility Service API to gain a system-level semantic understanding of UI elements. Accessibility API provides information about on-screen elements via AccessibilityNodeInfo objects~\cite{accessibilitynodeinfo}, encapsulating the data of UI components. 
    
    \tool retrieves interactive UI elements and classifies whether they are clickable, scrollable or editable. We use the extracted textual data from TextView to label interactive nodes, which are used for matching with the target from the users' command. For non-TextView interactive elements, \tool gains an understanding of them by investigating the effect of UI elements grouping~\cite{zhang2021screen}. Specifically, we obtain its label by searching for the adjacent TextView element which shares the same parent node, inspired by the label searching algorithm in Yang's previous work~\cite{yang2020dothishere}. We also extend the search range to retrieve textual information from the child components which is located inside the interactive element to improve the algorithm's coverage. \tool uses spatial information about UI elements to identify adjacent and overlapping elements, which provides contextual information to improve the semantic analysis of the UI. 

    \begin{figure}
    \centering
    \includegraphics[width=0.5\textwidth]{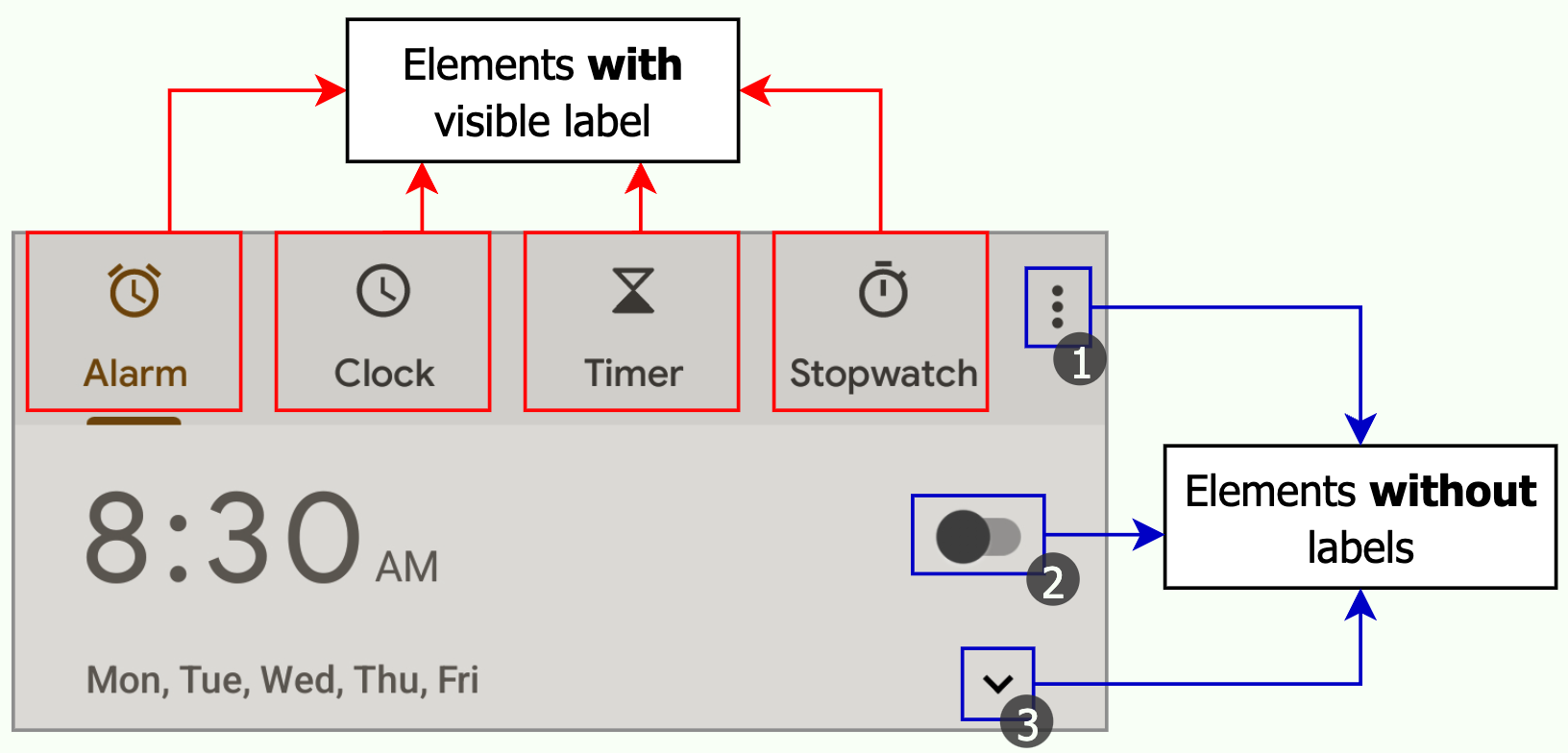}
    \caption{We show an example of a normal UI semantic where interactive elements are extracted. These elements can be invoked using the corresponding texts (elements with visible labels) or provided tooltip numbers (elements without labels).}
    \label{fig:UIsemantic}
    \end{figure}
    
    On several occasions, an interactive element (e.g., an icon) is not attached with any visible labels on the screen, hence the user is unable to specify the UI element by mentioning its label. Following the implementation of Voice Access~\cite{hume_2020}, we provide a tooltip labelling system, where a distinct number is dynamically assigned as a temporary label for unlabeled elements. By using the \emph{Layout Inflater}, we augment a numbering tooltip next to the unlabeled UI element by tracking its absolute coordinator on the screen. In this way, users can interact with unlabeled elements on the screen using the given number. For example, in Fig.~\ref{fig:UIsemantic}, the top right icon is appended number 1 since it does not provide any visible label, in which the user can tap it by saying \emph{``tap number 1''}. \tool also caters for users with special visual requirements by allowing runtime customization of tooltips' appearance, including the size, colour and opacity.

\subsection{Command Parser}
\label{sec:commandParser}
To understand user commands, most current voice assistant systems use various Natural Language Understanding (NLU) components to convert the command utterances into the formal meaning representation (MR)~\cite{louvan-magnini-2020-recent,li2020context}. Similarly, we proposed a semantic parser as our NLU component to convert utterances into formal meaning representations, which represent structured actions in our system. The generated MRs are then delivered to the Dialogue Manager (Section~\ref{sec:actionExecution}) system such that the module can understand and fulfil the user requirements. In this section, we will explain i) how we define the formal meaning representations, ii) the structures of our semantic parser and iii) how we generate training data for learning the semantic parser.
    \begin{figure}
    \centering
    \includegraphics[width=\textwidth]{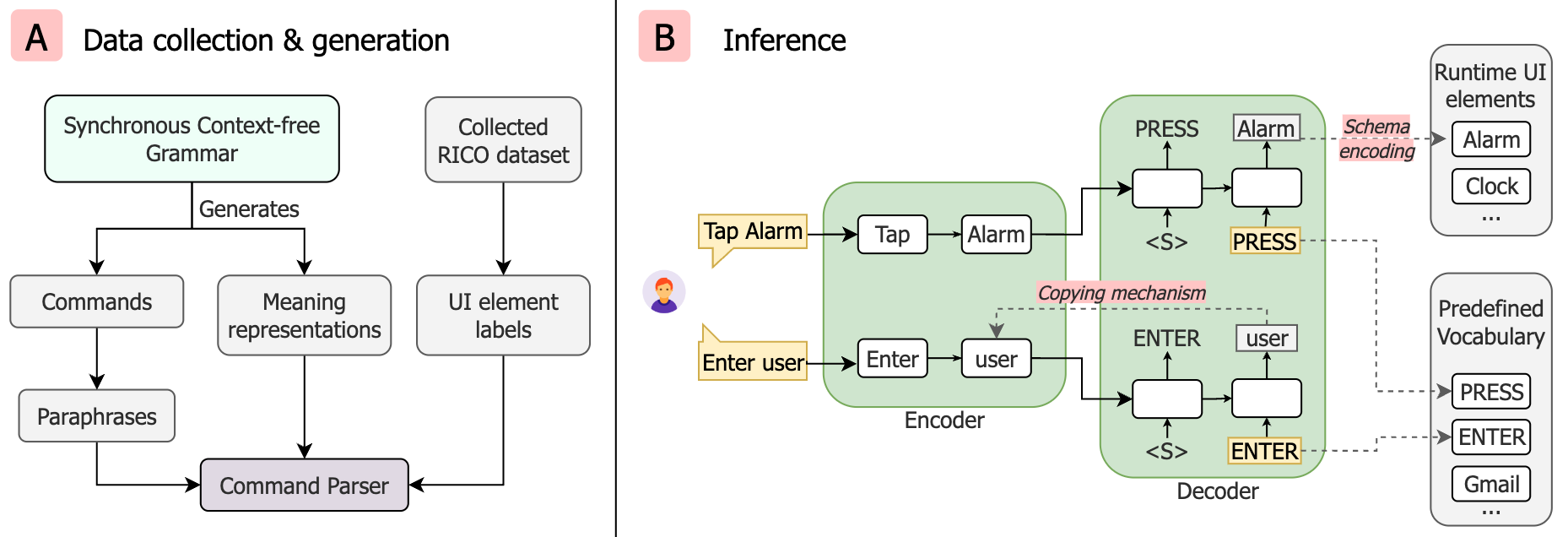}
    \caption{(A) Data collection process for training the command parser using the Overnight approach. (B) Example of the parser inference process for parsing human commands. }
    \label{fig:parser}
    \end{figure}
\subsubsection{Meaning Representation} 
Currently, the NLU system in the other voice assistants adopts frame-based meaning representations composed of intents and slots~\cite{gupta2018semantic,louvan-magnini-2020-recent}. However, such representation is known as having difficulties dealing with complex logic such as conjunction and negation in natural language~\cite{cohen2020back}. To avoid the restrictions of frame-based representations, we propose a novel MR language, \textit{VoicifyLang}. VoicifyLang defines several actions in the user commands, such as tapping, scrolling or entering text onto the UI elements. VoicifyLang also categorizes the action targets such as apps, components in the apps, the buttons on the screen, the input text and the directions of scrolling into different primitive types. 

We defined the language in a way that it is flexible to deal with more complex logic operations in the human language. Therefore, the language is highly extensible to support more features in \tool. The detailed syntax of our meaning representation is described by Abstract Syntax Description Language~\cite{wang1997zephyr}.
    
\subsubsection{Semantic Parsing}

Each MR in VoicifyLang comprises a sequence of tokens. We propose a novel parser, namely \textit{VoicifyParser}, to convert each user command into a sequence of MR tokens. The architecture of VoicifyParser is mainly inherited from BERT-LSTM~\cite{xu2020schema2qa}, a semantic parser which uses an attention-based Sequence-to-Sequence neural network~\cite{bahdanau2015neural} as the backbone. VoicifyParser builds a copying module and a schema encoding module onto the Sequence-to-Sequence to solve our task-specific problems, as illustrated in Fig.~\ref{fig:parser}B. Specifically, the encoder of the Sequence-to-Sequence is a pre-trained language model, BERT~\cite{liu2019roberta,kenton2019bert} and the decoder is a Long-short Term Memory~\cite{hochreiter1997long}. The vanilla Sequence-to-Sequence model only generates the tokens stored in a fixed vocabulary. Specifically, the vocabulary for VoicifyLang contains the tokens of actions, targets and some special tokens such as the delimiters and placeholders. The actions are pre-defined by VoicifyLang, while targets, including apps, components and buttons, are extracted using the Data Collection module defined in Section~\ref{sec:dataExtraction}. However, if the user command inputs the text, the text in the generated MR should all be copied from the user command utterance, given the VoicifyLang definition. Thus, the VoicifyParser employs a copying mechanism~\cite{gu2016incorporating} that allows the parser to copy the text from the user utterance. Another problem is that the buttons on each app page are usually dynamically generated, which means many buttons are out of vocabulary. To solve this problem, VoicifyParser adopts a schema encoding mechanism~\cite{wang2020rat}. When the input user command intends to tap a button, \tool collects all button names on the screen in runtime and sends them to the VoicifyParser. The schema encoding enables the VoicifyParser to generate tokens of a button out of the extracted buttons instead of only from the vocabulary.

\subsubsection{Data Synthesis}

The training data includes the pairs of user utterances, their aligned VoicifyLang MRs, and optionally a list of button names used to simulate the scenario in which the user wants to tap a button on a screen. Since manual annotation of user utterances is cost-intensive and time-consuming, we adopted a widely-used semi-automatic data collection method for semantic parsing, namely the Overnight approach~\cite{wang2015building}, as illustrated in Fig.~\ref{fig:parser}A. First, we manually wrote a set of synchronous context-free grammar (SCFG)~\cite{chiang2007hierarchical} rules. Expanding the SCFG rules could generate pairs of semantically equivalent canonical utterances and MRs. In the generated dataset, each MR has only one aligned canonical utterance. However, in the real-world scenario, natural language riches in linguistic variations. User speakers may vary word choices and morphologies for the same actions and targets in the MR and the syntactic structures for the utterances with the same MR. Therefore, we paraphrased each canonical utterance into multiple utterances with the same semantic meaning such that each MR would have multiple aligned utterances. The studies in~\cite{wang2015building,xu2020autoqa,shiri2022paraphrasing} validate that such paraphrases could significantly improve the performance of the semantic parser. We applied automatic paraphrasing methods to reduce the paraphrasing cost. We used the best-performing paraphrase method, a commercial online paraphrase service, Quillbot\footnote{https://quillbot.com/}, as in~\cite{shiri2022paraphrasing}. For each user command whose intention is to tap a button, we randomly sampled a list of buttons from the pre-defined set of buttons as the candidates for tapping. The button names are collected from the Rico dataset~\cite{deka2017rico}, which contains button names extracted from 10k Android UI screenshots. Thus, the parser can learn to generate tokens of the out-of-vocabulary buttons. The parser trained on the dataset generated by Overnight is proven to be robust to the richness and lexical and morphological diversities in the natural language~\cite{wang2015building}.

\subsection{Dialogue Manager}
\label{sec:actionExecution}

 Dialogue Manager extracts the pair of action-target from the received MR and performs relevance ranking to identify the suitable target from the collected data. Finally, \tool's executor performs the action on the user's device using Android-supported features.

    \subsubsection{Application Feature \& UI Element Matching}
     The matching between application components and user-requested component consists of two steps. First, the system searches for the application name from the meaning representation in the installed application namespace. After identifying the matched application, \tool searches for the component using the feature name and sends it to the executor. We search for the application name first to shrink the component search space and improve efficiency. 
    
    Since UI elements are dynamically changing when users are interacting with the phone, \tool must continuously capture and analyse new UI elements. \tool subscribes to the \emph{typeWindowContentChanged} accessibility events~\cite{accessibility_service} to get notified when changes happen to user's screen. The tool maintains separated dictionaries of labelled and unlabelled on-screen elements, which are updated using the data in \emph{AccessibilityNodeInfo} objects delivered by Android OS, as described in Section~\ref{sec:dataExtraction}. Upon receiving meaning representation, the matching module uses the attached UI element to get the corresponding node object, and deliver it to the executor to perform the action. 
    
    \subsubsection{Executor}
    Android Accessibility API provides a wide range of system control methods that allows accessibility service developer to have full control over users’ device. Thus, \tool used the \emph{performAction()} method~\cite{accessibilitynodeinfo} to perform actions, such as tapping, scrolling or entering text onto the UI elements. The executor sends the formed intent messaging object to directly open in-app components. On occasion, when none of the collected intents matches with the user's request, we open the application entry screen as a baseline.
    
    We implemented a queue containing action-target pairs to manage the sequential order of input actions. The executor will automatically pop the next action from the queue once the preceding action is successfully invoked, which allows command chaining to improve the usage efficiency. We implemented an automated validation system that verifies if the action is executed by monitoring the changes in the UI node metadata. Lastly, \tool sends audio feedback to users as a confirmation that the system is ready for the next command.

\subsection{Implementation}

 The Android client (Section~\ref{sec:dataExtraction} and Section~\ref{sec:actionExecution}) is implemented in Java using Android Studio. We utilized the native Google speech recognizer \cite{speechrecognizer} to interpret users' speech into the textual format and Google Text-to-speech engine~\cite{android_developers_2022} to provide audio feedback to users. We used Android Accessibility API \cite{accessibility_service} to implement a background service that can retrieve on-screen UI elements and perform actions on the screen. We implemented Python scripts to perform data collection and generation for training the parser. The back-end server that contains the semantic parser for analysing the user's command is developed using Flask framework \cite{grinberg2018flask} (Section ~\ref{sec:commandParser}). The communication between the back-end server and the Android client is handled using HTTP requests.   

\section{Technical Evaluation}
\label{sec:technicalevaluation}

In this section, we validate the performance of 2 novel components inside the system, namely i) the command parser and ii) the application feature retrieval module. We will first measure the accuracy of generating the meaning representation for natural human command. In addition, we perform a quantitative analysis of a set of application features in comparison with Google Assistant as the baseline.
\subsection{Command Parser Evaluation}

\subsubsection{Experiment Setup \& Metric}
We evaluate the parser with respect to its ability to convert the user command into the correct meaning representations. The parser is firstly trained on a synthetic generated dataset curated by the Overnight approach. Then the parser converts the commands in a manually collected test set into MRs. Then we compare the parser-generated MRs with the ground truth MRs and calculate the parser performance in three metrics.

 For the \textit{test set}, we first crawled user commands from the AndroidHowTo dataset~\cite{seq2act}, which contains step-by-step commands on achieving different tasks in Android. Then, we filtered out irrelevant data entries such as commands to interact with the other operating systems to obtain 101 clear commands for the test set. Each test case is a single command to perform the step, such as ``Tap Clear browsing data'' or ``Go to the Profiles tab''. The parser requires a list of current on-screen elements for each tapping interaction. Therefore, we extracted the list of clickable elements for each tapping command from the Rico dataset~\cite{deka2017rico}, which contains the metadata to identify clickable elements in each screen from 10k Android apps. Finally, the ground truth MR of each test case is manually annotated by one student and validated by another to ensure the annotation correctness. Such a test set includes rich linguistic variations. For example, given our manual calculation, one MR has an average of 1.2 aligned natural language commands. The actions such as \textit{OPEN}, \textit{PRESS}, \textit{SWIPE} and \textit{ENTER} are described by 4, 6, 2, and 2 different phrases, respectively. Evaluating our parser on this test set can validate our parser's robustness to the rich linguistic variations.


To evaluate the performance of the parsers, we adopt three evaluation metrics, \textit{Exact Match Accuracy} (EM Accuracy)~\cite{dong2018coarse,li2021few}, \textit{Target F1} and \textit{Action F1}. Exact Match Accuracy is one of the most commonly used metrics in semantic parsing, which computes the percentage of the commands which are correctly mapped to its ground truth meaning representations. To further understand how well the parser could predict the correct actions and targets, we also computed the F1 scores of the targets and actions. The \textit{Target F1} and \textit{Action F1} report the average F1 score of all targets and actions, respectively, weighted by their corresponding frequencies in the ground truths. 

We compared with two baselines, vanilla Seq2Seq~\cite{bahdanau2015neural} and BERT-LSTM~\cite{xu2020autoqa}. Vanilla Seq2Seq includes an LSTM encoder and an LSTM decoder. BERT-LSTM replaces the LSTM encoder of Seq2Seq with a pre-trained language model, BERT~\cite{kenton2019bert}, and an additional copying module on the LSTM decoder. As our model is specifically designed to solve the issues of the out-of-vocabulary on-screen buttons, we also evaluated the performance of BERT-LSTM and VoicifyParser when all the buttons are not stored in the pre-defined vocabulary list. As such, the parser can only access the button names from the set of extracted on-screen buttons in runtime.

\subsubsection{Evaluation Result}
As in Table ~\ref{table:parser}, our parser, VoicifyParser, can achieve an EM accuracy as high as 90\% on the test set with rich linguistic variations. VoicifyParser also consistently outperforms the other two baselines in generating complete MRs or the individual actions and targets in the MRs. Compared with the vanilla Seq2Seq, the accuracy of VoicifyParser is even 40\% higher. Due to a lack of a copying mechanism, Seq2Seq can not correctly parse most of the input-text commands. For example, \textit{enter `4-digit PIN code' .} is incorrectly parsed into \textit{( ENTER , ` 4-digit PIN ' )}. In our experiment, it fails on all such types of questions in the test set, which occupies 5\% of the total test set. Vanilla Seq2Seq employs no pre-trained language model, so it is not robust to the rich linguistic variations in the user commands. As a result, it performs significantly worse than BERT-LSTM and VoicifyParser in all three aspects. BERT-LSTM and VoicifyParser have comparable performance in terms of Action F1. However, BERT-LSTM lacks a schema encoding mechanism as our model. Thus, it can not utilize the runtime on-screen information. Even with all the buttons stored in the pre-defined vocabulary list, the Target F1 of BERT-LSTM is still 7\% lower than VoicifyParser, depicting that on-screen information is a great boost to the parser's ability with respect to predicting the targets if the parser employs the schema encoding. With all the button elements removed from the vocabulary, the performance of BERT-LSTM drops significantly. BERT-LSTM (OOV) performs even worse than the Seq2Seq parser in all three metrics. Surprisingly, removing the button targets also drops the action prediction performance for the BERT-LSTM (OOV). For our VoicifyParser (OOV), the influence of removing button names is negligible, showing that schema encoding improves the robustness of our model to the out-of-vocabulary buttons in the test commands.

\begin{table}[]
\caption{The experiment results of different parsers in terms of three metrics. \textit{OOV} (Out of Vocabulary) indicates that the button names in the ground truths of the test set are not stored in the vocabulary of the parsers. }
\begin{tabular}{cccc}
\hline
\textbf{Parsers} & \textbf{EM Accuracy} & \textbf{Target F1} & \textbf{Action F1}\\ \toprule
\textit{Seq2Seq} & 52.48 & 51.33 & 83.17  \\ \hline
\textit{BERT-LSTM} & 80.20 & 85.71  & 96.55  \\ 
\textit{BERT-LSTM (OOV)} & 29.70 & 35.41  & 60.27  \\ \hline
\textit{VoicifyParser} & \textbf{91.09} & \textbf{93.05}  & 97.03  \\ 
\textit{VoicifyParser (OOV)} & 89.11 & 90.76  & \textbf{98.02}  \\ \hline
\end{tabular}

\label{table:parser}
\end{table}
\subsection{Application Feature Coverage}

    \begin{figure}[!h]
    \centering
    \includegraphics[width=\linewidth]{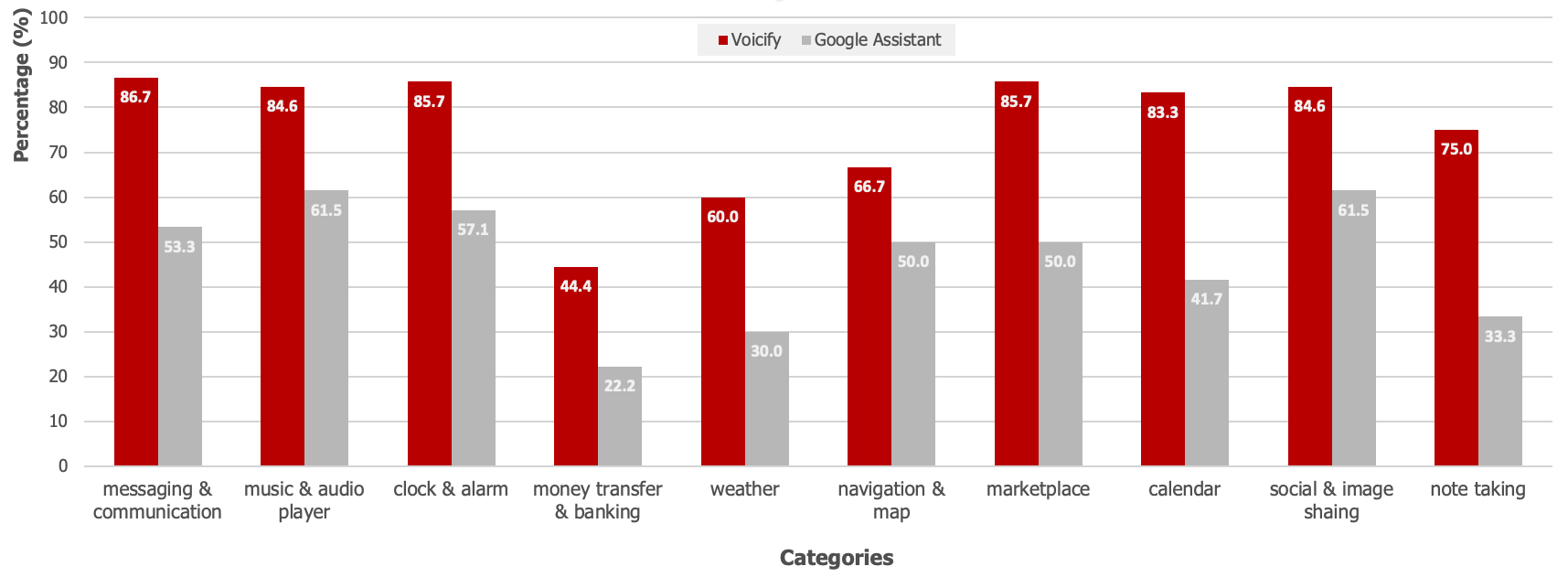}
    \caption{The comparison of feature coverage between \tool and Google Assistant across 10 different app categories.}
    \label{fig:coverage}
    \end{figure}

    \subsubsection{Experimental Setup \& Metric}
    
    In this experiment, we validate the ability of \tool to directly open the installed application's features.We consider an in-app feature as a specific screen that serves the functionality to users. Given a set of features from various installed apps, we measure the application feature coverage by manually opening each feature via voice command and recording the success rate. Since Voice Access does not support opening features from other apps using voice commands, we used Google Assistant~\cite{google_assistant} as the baseline. Using the result from the formative study that investigated common user tasks on smartphone~\cite{arsan2021app}, we identified the 10 most common categories that will be evaluated in this experiment. For each application category, we randomly picked 3 applications from the Google Play Store top suggestions. All chosen applications are popular, ranging from 1 million downloads to more than 1 billion downloads. For each application, we identify a set of provided features that the application developer stated in the Play Store listing description. The number of features from each app varies, ranging from 2 to 5 depending on the complexity of the app. For example, applications in \emph{Clock \& Alarm} category are simple, only allowing users to add new alarm and view the list of added alarms, while applications in \emph{Post a picture} category like Instagram offers many more features. In the end, we collected 117 test cases from 30 apps, each of them being an important feature stated by the developer when publishing the app on Play Store. A full list of the apps and features can be found at the Github repository\footnote{\url{https://github.com/vuminhduc796/Voicify}}.

     For each test case, we first record the ground truth as the screen that can provide the requested functionality to users. We used the success rate as the primary metric, each test case will be marked as successful if the tool can directly open the desired screen, otherwise, it will be marked as failed. We compared the success rate of \tool and Google Assistant to validate the performance of the direct invocation module.

    \subsubsection{Results}
     Fig.~\ref{fig:coverage} compares the feature coverage by category between \tool and Google Assistant. Overall, \tool successfully located and invoked 90 application features out of 119 test cases (76.9\%), compared to 55 features from Google Assistant (47.0\%). Both tools have achieved high performance in popular applications such as Yahoo Mail, SoundCloud and Twitter with over 80\% success rate. However, the feature coverage from Google Assistant dropped dramatically for less well-known apps because those applications are missing the required declaration from their developers to integrate with Google Assistant. Since our approach does not require any developer efforts to integrate the application with \tool, we achieved a solid performance across a wider range of apps.
    
    We performed the error analysis by investigating failed test cases to understand the main problems that affect \tool feature coverage. First, as mentioned before, some important features are encapsulated inside the application, which does not allow direct access from other applications. Hence, both \tool and the baseline have poor performance for \emph{money transfer \& banking} apps as those applications might have enhanced protection, blocking direct access to certain features to prevent malicious attacks. Second, some application screens require data that is passed from the previous screen. This data is often stored in a bundle, attached to the intent to open. For example, when opening the video player from YouTube, the attached data must specify which video will be played using a certain video ID. \tool is unable to identify and populate the required data, hence failing to directly open those features. Third, some applications have implemented a centralized activity as the entry point for all external invocations through intent. This implementation limited the list of features that is can be captured by \tool, hence impacting the feature exploration of \tool.

\section{User Study}
\label{sec:evaluation}
To demonstrate the usefulness of our tool in practice, we conducted a user study to evaluate \tool system as a whole in real-world scenarios. As a baseline, we compare \tool with Voice Access since the tool is currently the best voice command available in Android~\cite{yamada_2020}. The goal of the study is to i) benchmark the user performance with \tool, compared to the baseline and ii) compare the user feedback on the cognitive load and usability of \tool to the baseline and lastly iii) collect qualitative feedback to suggest future improvement on \tool. We record the time taken to finish each task using \tool and Voice Access. In addition, we conducted a post-experiment interview with each user to collect both quantitative and qualitative feedback. 

\begin{table}[]
\caption{The list of tasks for user evaluation.}
\begin{tabular}{cllc}
\hline
\textbf{No} & \multicolumn{1}{c}{\textbf{Task}} & \multicolumn{1}{c}{\textbf{Application}} & \textbf{\#Steps} \\ \hline

Task 1 & Check for a saved cooking recipe. & World Cuisines & 6 \\ \hline
Task 2 & Check for cooking steak instructions and set a timer. & Steak Timer & 8 \\ \hline
Task 3 & Convert the mass of ingredients from teaspoon to tablespoon. & Unit Converter Ultimate & 10 \\ \hline
Task 4 & Add a grocery shopping note. & Fast Notepad & 12 \\ \hline

\end{tabular}

\label{table:tasks}
\end{table}

\subsection{Tasks}

Based on an example scenario of having both hands busy when cooking as provided by participants in Section~\ref{sec:study}, we created 4 relevant tasks that users would perform on mobile devices when cooking. The tasks covered most common interactions on the screen, including tapping, swiping and entering text. We sorted the tasks based on the number of steps to achieve the task using voice command, hence from task 1 to task 4, the difficulty level increased. The list of tasks is mentioned in Table~\ref{table:tasks}. To introduce the task to participants, we provided written step-by-step tutorials on how to achieve the task. We also recorded a walk-through video for each app to demonstrate the tasks using physical touches and the expected outcome.

\subsection{Participants}

We recruited 8 participants (6 males and 2 females) who can speak English at a proficient level. In addition, all participants have a good level of familiarity with technological devices, as they all use a smartphone regularly. Although participants had relevant exposure to virtual assistants such as Siri or Google Assistant, none of the participants were familiar with using assistive tools to control their smartphones with voice commands, especially none of the participants used any of the experiment tools. The reason for selecting this set of participants is that in this study, we also measure the learnability of the experimental tools. Each participant is awarded an A\$50 gift card for their participation.

\subsection{Procedure}
We invited each participant to join a face-to-face meeting individually for the user evaluation. We set up one experimental device for all experiments, which is a Google Pixel 5 (Android 11) since some users do not have an Android device. In addition, some applications are not installed on participants’ devices or require a registered account to proceed. Lastly, different devices might have different response rates, hence affecting the correctness of results.

First, we provided users with some basic understanding of Android OS, and we introduced Google Voice Access and \tool. We recorded demo videos on how to achieve tasks with \tool. Since Voice Access has included a built-in tutorial for new users, we provided that material to the participant. After that, we trained users to achieve some basic tasks using both applications and allowed the users to practise using each tool for equally 5 minutes. We also introduced the applications that will be used for the evaluation. Since the majority of users were unfamiliar with the tasks and experimental applications, we guided them through the steps for each task and let them achieve each task with physical interactions to memorise the task. In the end, each user confirmed to have an approximate level of understanding of Android OS, both voice control applications and the experimental tasks. 

After the training, we asked each participant to complete 4 different tasks with no interventions by the experimenters. Participants used \tool to complete 2 tasks and used the baseline tool to complete 2 other tasks. They are not aware of which tool is developed by us. The order of the tasks and used tools will be rotated for each participant in a counter-balanced manner~\cite{depuy2014counterbalancing} to avoid potential biases. For example, P1 first completed task 1 and task 4 using \tool and then completed tasks 2 and task 3 using Voice Access while P5 completed task 1 and task 2 using Voice Access before completing task 3 and task 4 using \tool. For each given step, we had a cut-off time of 60 seconds if the participant could not figure out the way to complete the step using the voice command.

We recorded the time taken to fulfil each task, including the cut-off time to perform quantitative analysis. We collected 32 data entries since each of the 8 participants has finished 4 tasks. In the end, using the System Usability Scale (SUS)~\cite{brooke1996sus} form with a 5-point Likert scale, we evaluate the usability of \tool, compared to Google Voice Access. In addition, we investigated the cognitive load when experimenting with each tool using the NASA-TLX~\cite{hart2006nasa} form with a 7-point Likert scale. Lastly, we collected qualitative feedback on which part they liked the most about \tool and what might improve the system.

\subsection{Result}
\label{sec:result}
    \begin{figure}
    \centering
    \includegraphics[width=\linewidth]{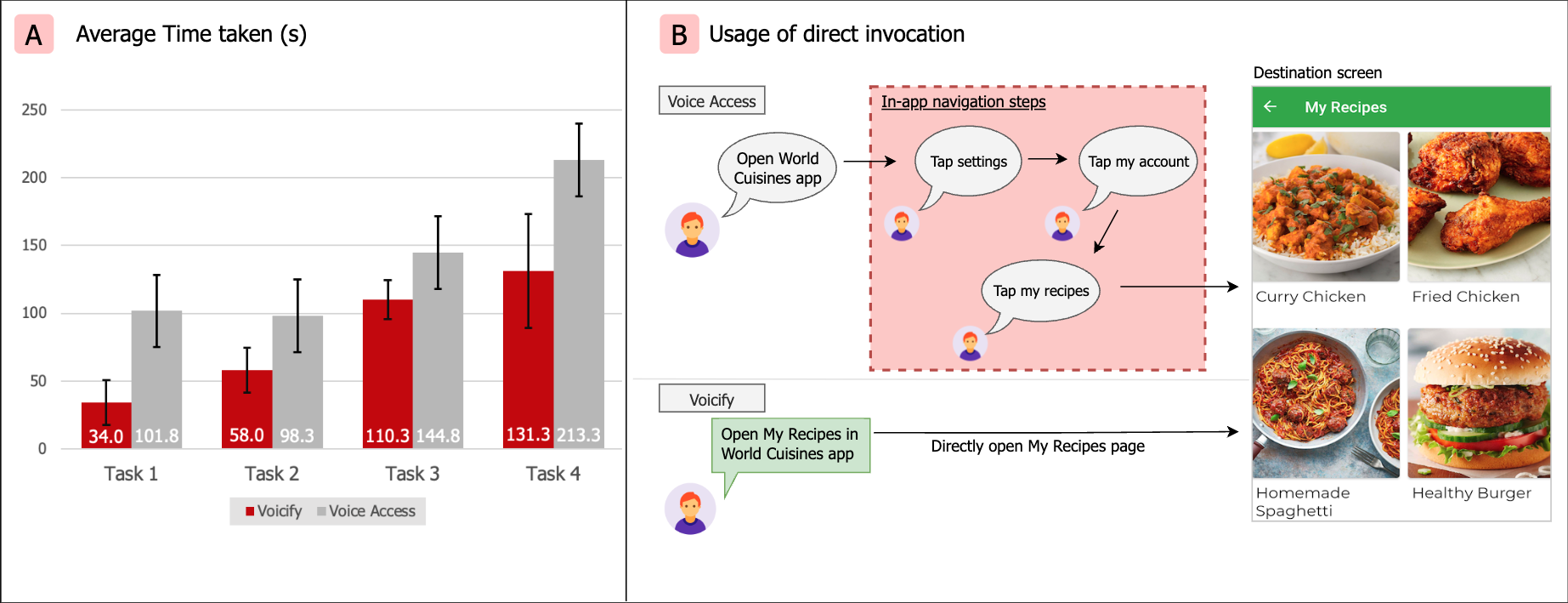}
    \caption{(A) Efficiency comparison between \tool and Voice Access. (B) The usage of direct invocation with \tool in task 1.}
    \label{fig:data}
    \end{figure}
    
    \subsubsection{Overall User Performance}
    We performed a quantitative analysis of the time taken (in seconds) to complete each task, as shown in Fig.~\ref{fig:data}(A). In general, as the number of steps increase, participants required more time to complete the task. The average time taken to complete the tasks with \tool is 93.2 seconds, compared to 140 seconds using Voice Access, resulting in a 33.4\% efficiency boost. We observed a significant disparity in recorded time due to (i) the usage of direct feature invocation in task 1 with \tool and (ii) the complexity in inputting text and tapping unlabelled icons with Voice Access in task 4. Specifically, participants directly opened the list of saved recipes in task 1 using \tool, hence they skipped 3 steps as described in Fig.~\ref{fig:data}(B), resulting in a shorter time. For task 4, participants were requested to input text into several text boxes, which caused some issues in selecting the text box and typing the text. In addition, some steps in task 4 required users to tap an unlabelled icon (e.g., the star icon to add a note to the favourite list), therefore, users made several unsuccessful attempts to guess the corresponding label of the icon. Using the grid-based tapping from Voice Access was proven to be inefficient as it required extra commands from users to show/hide and change the granularity of the grid. \tool solved the problems by attaching a numbering tooltip next to the unlabelled icon, allowing users to promptly perform interactions. By observing the experiment, we recognized the issue in transforming the human voice into the textual format. \tool did better in post-processing the raw user command using predefined heuristics to resolve common transcription errors. However, we noticed that \tool requires slightly more time to process each command because of the latency in API communication to the backend server. This latency caused a short period of unresponsiveness after recording the user command, adversely affecting the performance and user experiences. The problem could be mitigated in the future by deploying a mobile version of the model locally on user devices.
    
    \begin{figure}[!h]
    \centering
    \includegraphics[width=\linewidth]{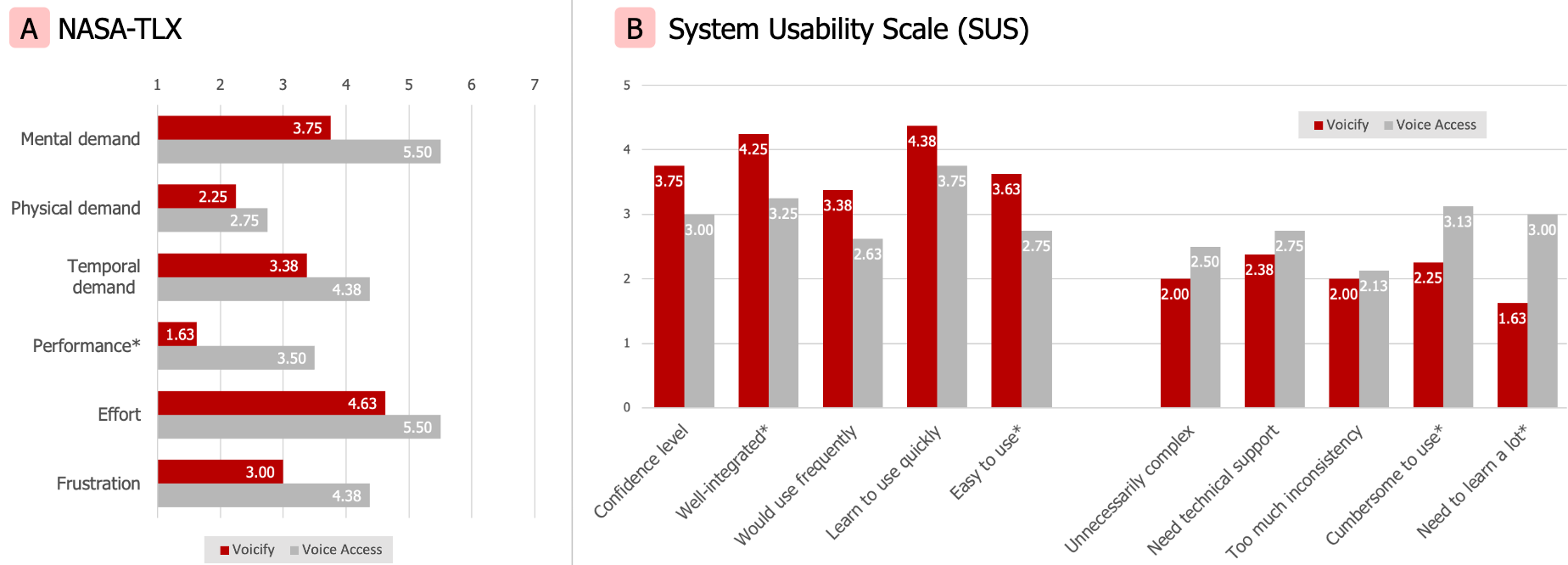}
    \caption{The comparison between \tool and Voice Access for A) the average cognitive load when using NASA-TLX form (lower is better) and B) the System Usability Scale (SUS). *: statistically significant (p < 0.05).}
    \label{fig:chart}
    \end{figure}
        
    \subsubsection{Cognitive Load \& Usability Ratings}
    Fig.~\ref{fig:chart}(A) summarizes the participant's feedback on their level of cognitive load for each system using the NASA-TLX form. With a lower level of effort, participants achieved significant improvement in the performance using \tool (t = -2.61, p = 0.035) as a result of direct invocation usages. In addition, the command parser was precise in mapping faulty commands to correct actions, helping users to easily interact with the system. Participants experienced less frustration using \tool, as well as confirmed that \tool required less mental demand, compared to Voice Access. The result proved a significant improvement of \tool, fulfilling the design implication of reducing the required cognitive loads to operate the tool.

    We compared the quantitative feedback of participants, including 10 design and usability questions on a 5-point Likert scale and applied a pairwise t-test to the result, as shown in Fig.~\ref{fig:chart}(B). The result validated that we improved the usability of the voice controlling system, as the average SUS score for \tool is 72.813, while Voice Access received 54.688.
    \tool was better integrated than Voice Access (t = 2.65, p = 0.033). Due to the intuitiveness of the tooltip labelling system and direct invocation, \tool received significantly better ratings across multiple indicators. The system is confirmed to be less cumbersome to use (t = -2.45, p = 0.041) and required less learning (t = -2.58, p = 0.036), compared to the baseline. Participants confirmed an improvement in the ease of operation (t = 1.99, p = 0.041) since tooltip selection was more convenient compared to the grid-based selection from Voice Access to tap unlabelled icons.
    
    \subsubsection{Qualitative Feedback}
    In this section, we collate qualitative feedback from participants after experimenting \tool and Voice Access. Overall, the participants are satisfied with the tool, as well as providing suggestions for further improvements.
     
    \textit{Innovative method of interacting with phone devices.}
    Participants who have never used a voice control system to perform a sequence of daily life actions were eager to explore it further. P5 expressed that \emph{``I really enjoy using the voice to control the phone, it is a new concept to me"}. Moreover, participants who are tech savvy are impressed by the capacity of voice assistant technology, as evident in P6’s comment that \tool \emph{``can do most of the things that I need"} and \emph{``understand what I want to say"}. In normal circumstances, participants prefer the traditional tapping interaction over voice command despite their positive experience with \tool. It is then concluded that even though voice command technology is excitingly novel and capable, the barrier to its wide adoption is the intricate human behaviour adaptation process.

    \textit{Usages of direct invocation.}
    \tool’s novel feature of direct invocation was implemented in the experiment and suggested as an option to carry out a sequence of activities, which garnered positive feedback from P1 and P7. Participants mentioned that the direct invocation feature was a great advance for voice control systems, as it allows users to \emph{``quickly access"} the desired screen in a particular application, alleviating the complexity of the task by reducing the number of steps and wait time. Therefore, future improvement will include expanding the set of in-app components that can directly be opened from the main screen using \tool.

    \textit{Numbering tooltips as an effective solution.}
     \tool’s tooltip labelling system received overall positive feedback. P2 mentioned that the best thing about \tool was \emph{``the ability to identify icons in an app in numbers whereas in Voice Access one may need to use grid selections"} and P4 appreciated the bespoke ability to tap on \emph{``labelling icons for which you may not know the name/label using numbers"}. P3, P5 and P7 agreed that numeric labelling has helped them to quickly interact with different icons on the screen without having to know the name of the icons, which is convenient and stress-free. In contrast, participants noted the demerit of the grid-based selection from Voice Access. P1 and P3 expressed that the grid-based selection is \emph{``difficult''} and \emph{``imprecise''}. The result showed the inefficacy of grid-based selection from Voice Access and the user preferences towards tooltips selection from Voicify.
    
    \textit{UI design improvement \& system feedback.} Participants acknowledge the great UI design and responsiveness of Voice Access while giving Voicify suggestions to deliver better user experiences. P3 mentioned Voice Access's \emph{``impressive capability to interpret the speech on the go''} and P7 expressed great interest in the real-time \emph{``closed-caption''} that Voice Access generated. For \tool, P2 and P6 mentioned that the toggle button to start the tool and the system status indicator blocked certain UI components. To tackle this issue, possible solutions include moving overlaying components onto the notification bar or providing the ability to show or hide UI components of \tool in runtime using a predefined voice command. P7 has suggested that displaying \emph{``closed-caption as we talk to give a sense of immediate feedback to the user"}, which will improve the responsiveness of the \tool.

\section{Implications and future works}
\label{sec:discussion}

\tool’s capability and a seamless user experience are one of our priorities for further development which will propel mainstream adoption of voice interaction for multiple ubiquitous computing devices. In this section, we will discuss the implications and propose further improvements to the current system.

\subsection{Interacting with Unlabelled Icons and Images}
Our study suggested the applicability of tooltips to improve the efficiency and accuracy of voice interactions for multiple devices, such as smart cars and smart TVs. It helped to overcome shortcomings of extant voice-based selection methods, especially when it is critical to interact with low mental demand. From the experiment result in Section~\ref{sec:result}, tooltip labelling significantly improved user performance as it was more precise than grid-based selection with less visual occlusion. Using the tooltip system from \tool, users were able to directly mention the number without the need to search for the matched tile in the grid selection system. As result, tooltips helped reduce the required cognitive effort to operate the system measured using the NASA-TLX form. During the experiment, when applying grid-based selection for screens with clustered elements, multiple elements fitted in one tile due to the low granularity of the grid, causing unwanted interactions. Participants expressed frustration when misclicking an element, causing extra navigation steps to finish the task. In addition, the tooltip system helped improve the learnability aspect of the system, as fewer steps and simpler command syntax are needed to perform the same interaction.

\subsection{Mapping User Commands}
In this work, we proposed an advanced deep-learning parser that interprets human commands to produce structured actions. The parser is designed to understand the nuance in human language and suggest the closest match for user queries. The result from Section~\ref{sec:technicalevaluation} showed that VoicifyParser outperformed other advanced parsers due to its capability to handle usages of synonymous words and out-of-vocabulary labels. The ability to interpret and record new out-of-vocabulary words allows the vocabulary size to grow in runtime, in results allowing the parser to work seamlessly in real-world scenarios. The user study has proven the effectiveness of our parser, which results in an improvement in user performance and a reduction in the required cognitive load to operate the tool, as shown in Section~\ref{sec:result}.

Although VoicifyParser is currently fine-tuned to interpret a fixed set of functionalities proposed by \tool, the parser is designed for high extensibility and usability. By providing additional training data for transfer learning progress~\cite{ezen2020comparison}, the model can be extended to cater for other meaning extraction problems with minimal efforts, such as controlling household appliances and smart vehicles. We hope to make the latest NLP technology more accessible and applicable for further research in the human-computer interaction domain. 

\subsection{Improvement for Feature Shortcuts}

 The findings imply the significance of providing shortcuts for users to achieve certain tasks, as performance is the top priority that determines user experience. In-app shortcuts helped users to achieve Task 1 in Section~\ref{sec:evaluation} with only one-third of the time taken without direct invocation, as well as receiving very positive feedback from user responses. Not only users who are unable to perform physical touch on the screen will be rewarded by the feature, but also users who can physically control the smartphone using their hands will use this feature to accelerate their tasks.

From the experiment, we observed that explorability and learnability are the key factors that affect the usability of feature shortcuts. When experimenting 
with \tool to achieve Task 1 which contains an in-app shortcut to view all saved recipes, a participant did not use the shortcut as she did not know if the shortcut was available and the feature name attached to it. We acknowledge the explorability and learnability issues of the feature and propose a better recommendation system with additional GUIs to introduce available shortcuts to the user. 

On the other hand, we acknowledged that most unexplored app features were encapsulated within fragments. The feature retrieval module in \tool is currently based on the content within the Manifest file in which all activities are declared. Nevertheless, this requirement does not apply to fragments, and hence, features located within these fragments remain undiscovered. According to Li et al., more than 50\% of the most popular apps are using Fragment as the basic building block~\cite{li2017data}. Therefore, the full retrieval and invocation of fragments are critical to improving \tool's exploration of all in-app components. Further studies on reconstructing fragments and fragment navigation within an activity are proposed to improve the granularity of feature exploration and provide forward compatibility for newer app versions.

\subsection{Limitation \& Future Works}
While invoking developer-defined deep links significantly improved our direct invocation module's coverage, the module will depend on the developer's implementation of deep links. Therefore, impaired deep links provided by developers may also affect the coverage of the direct invocation module from \tool. By default, we show users the app's launcher page if the deep link is broken. In addition, as developers may modify introduced deep links through app updates, the availability of deep links could be affected. \tool mitigated the problem by automatically updating the list of available deep links for installed apps, as deprecated deep links will be removed from the database and newly introduced deep links will be added. We also provide a GUI component within the \tool application that displays available deep links for each application to inform users about the availability of deep links. Lastly, as our tool depends on the integrated deep links, apps with fewer deep links will achieve lower feature coverage.

Another limitation to consider is that our selection of participants for user evaluation may have introduced validity threats. We chose participants who were not familiar with using voice commands to control their smartphones in order to minimize biases and assess the ease of learning for our system. Therefore, we acknowledge the absence of experimenting the tools with experienced users, which may provide other perspectives regarding the performance of each tool. However, we mitigated the stated issue by designing the experiment which is suitable for users with minor experience. Firstly, we first trained participants using both of the tools before conducting any experiments, which helped improve the user’s familiarity with voice commands. Secondly, the experimental tasks did not require an in-depth understanding of voice assistants and participants can fully achieve the tasks using the commands that they are trained with. Thirdly, while the level of experience from each user may affect the total time taken to finish the task, it does not affect the comparison between the tools as each same participant is asked to use both tools.

We propose several directions that \tool can be improved. Firstly, the current system requires users to maintain visual contact with the touchscreen, which is burdensome in use cases where the mobile device is out of sight. Therefore, we propose future work to integrate with screen readers such as Talkback in Android OS to cater for non-visual usages. Lastly, while \tool put forth an exciting opportunity to cater for motor impairments, we have not yet investigated the impact it has on users with physical constraints. Therefore, with further empirical research to improve voice assistive platforms for impairments, incremental improvements could be proposed on top of \tool to benefit different types of disabilities.

\section{Conclusion}
\label{sec:conclusion}

In this paper, we present the \tool system that enhances the usage of voice commands in Android devices. We proposed a novel parser that generates structured actions (also known as MR) for human commands to interact with the smartphone. The dialogue manager uses the actions to perform matching with the collected data and execute the actions on the device. \tool demonstrates a novel approach to utilise the on-device data including applications code base and on-screen UI semantics, therefore, the computation workload is lightweight and can be executed locally. From the experiment, the natural language parser achieves outstanding accuracy for the human command dataset, compared to multiple baselines. Our experiment also shows that \tool has an effective direct invocation module that has high coverage of application features without requiring extra developer effort to comply. We also conducted a user evaluation that indicates the high usability of the system in real-world tasks, fulfilling all design implications. Lastly, since \tool is an open-source project, it lays the ground for future works on the intuitiveness of user verbal input which contributes to the further improvement of human-computer interactions.

\bibliographystyle{ACM-Reference-Format}
\balance
\bibliography{sample}

\end{document}